\documentclass[aps,prb,twocolumn,showpacs,superscriptaddress]{revtex4}

\usepackage{graphicx}
\usepackage{amsmath}
\usepackage{multirow}
\usepackage{tabularx}

\usepackage[colorlinks,linkcolor=blue,anchorcolor=blue,citecolor=blue]{hyperref}

\begin{document}

\title{Evidence for charge and spin order in single crystals of La$_3$Ni$_2$O$_7$ and La$_3$Ni$_2$O$_6$}	
\author{Zengjia Liu}
\author{Hualei Sun}
\email{sunhlei@mail.sysu.edu.cn}
\author{Mengwu Huo}
\affiliation{Center for Neutron Science and Technology, Guangdong Provincial Key Laboratory of Magnetoelectric Physics and Devices, School of Physics, Sun Yat-Sen University, Guangzhou, 510275, China }
\author{Xiaoyan Ma}
\affiliation{Beijing National Laboratory for Condensed Matter Physics, Institute of Physics, Chinese Academy of Sciences, Beijing 100190, China }
\affiliation{School of Physical Sciences, University of Chinese Academy of Sciences, Beijing 100190, China}
\author{Yi Ji}
\author{Enkui Yi}
\author{Lisi Li}
\author{Hui Liu}
\author{Jia Yu}
\affiliation{Center for Neutron Science and Technology, Guangdong Provincial Key Laboratory of Magnetoelectric Physics and Devices, School of Physics, Sun Yat-Sen University, Guangzhou, 510275, China }
\author{Ziyou Zhang}
\affiliation{Center for High Pressure Science and Technology Advanced Research, Shanghai 201203, China}
\author{Zhiqiang Chen}
\affiliation{Center for High Pressure Science and Technology Advanced Research, Shanghai 201203, China}
\author{Feixiang Liang}
\affiliation{Center for Neutron Science and Technology, Guangdong Provincial Key Laboratory of Magnetoelectric Physics and Devices, School of Physics, Sun Yat-Sen University, Guangzhou, 510275, China }
\author{Hongliang Dong}
\affiliation{Center for High Pressure Science and Technology Advanced Research, Shanghai 201203, China}
\author{Hanjie Guo}
\affiliation{Songshan Lake Materials Laboratory, Dongguan, Guangdong, 523808, China}
\author{Dingyong Zhong}
\author{Bing Shen}
\affiliation{Center for Neutron Science and Technology, Guangdong Provincial Key Laboratory of Magnetoelectric Physics and Devices, School of Physics, Sun Yat-Sen University, Guangzhou, 510275, China }
\author{Shiliang Li}
\affiliation{Beijing National Laboratory for Condensed Matter Physics, Institute of Physics, Chinese Academy of Sciences, Beijing 100190, China }
\affiliation{School of Physical Sciences, University of Chinese Academy of Sciences, Beijing 100190, China}
\affiliation{Songshan Lake Materials Laboratory, Dongguan, Guangdong, 523808, China}
\author{Meng Wang}
\email{wangmeng5@mail.sysu.edu.cn}
\affiliation{Center for Neutron Science and Technology, Guangdong Provincial Key Laboratory of Magnetoelectric Physics and Devices, School of Physics, Sun Yat-Sen University, Guangzhou, 510275, China }

\begin{abstract}
Charge and spin order is intimately related to superconductivity in copper oxide superconductors. To elucidate the competing orders in various nickel oxide compounds are crucial given the fact that superconductivity has been discovered in Nd$_{0.8}$Sr$_{0.2}$NiO$_2$ films. Herein, we report structural, electronic transport, magnetic, and thermodynamic characterizations on single crystals of La$_3$Ni$_2$O$_7$ and La$_3$Ni$_2$O$_6$. La$_3$Ni$_2$O$_7$ is metallic with mixed Ni$^{2+}$ and Ni$^{3+}$ valent states. Resistivity measurements yield two transition-like kinks at $\sim$110 and 153 K, respectively. The kink at 153 K is further revealed from magnetization and specific heat measurements, indicative of the formation of charge and spin order. La$_3$Ni$_2$O$_6$ single crystals obtained from topochemical reduction of La$_3$Ni$_2$O$_7$ are insulating and show an anomaly at $\sim$176 K on magnetic susceptibility. The transition-like behaviors of La$_3$Ni$_2$O$_7$ and La$_3$Ni$_2$O$_6$ are analogous to the charge and spin order observed in La$_4$Ni$_3$O$_{10}$ and La$_4$Ni$_3$O$_8$, suggesting charge and spin order is a common feature in the ternary La-Ni-O system with mixed-valent states of nickel.

\end{abstract}

\maketitle

\section{Introduction}

The recent discovery of superconductivity in Nd$_{0.8}$Sr$_{0.2}$NiO$_2$ films has aroused great research enthusiasm in the search for superconductivity and understanding the pairing mechanism in nickel oxide system\cite{li2019superconductivity}. Particularly, whether the mechanisms of the superconductivity in nickelates and copper oxide superconductors are the same remaining an open question. The superconductivity, spin order, and charge order could be tuned by carrier doping in cuprates. It is widely accepted that the static and dynamic spin and charge orders play a crucial role in the mechanism of superconductivity\cite{vignolle2007two,Keimer2015}.

The ternary nickel oxide $Ln$-Ni-O system ($Ln$ = La, Pr, Nd, Sm, and Eu) contains the Ruddlesden-Popper (RP) compounds $Ln_{n+1}$Ni$_n$O$_{3n+1}$, which possess $n$ layers of perovskite-type $Ln$NiO$_3$, separated by single rocksalt $Ln$O layer along the $c$ axis\cite{zhang2020self,zhang2020similarities,wang2020synthesis,lee2004infinite}. The Ni ions exhibit mixed valences of Ni$^{3+}$ (3$d^7$) and Ni$^{2+}$ (3$d^8$). By a topochemical reduction method, two apical oxygen atoms of a NiO$_6$ octahedra could be removed and the remained oxygen atoms rearrange and result in a fluorite $Ln$-O$_2$-$Ln$ layer, as shown in Fig. \ref{fig1}.  The topochemical reduced compounds $Ln_{n+1}$Ni$_n$O$_{2n+2}$ with mixed-valent states of Ni$^{1+}$ (3$d^9$) and Ni$^{2+}$ would form. The structures of the RP system and the PR reduced system are analogous to the ternary $Ln$-Cu-O system, especially the Ni-O planes that are regarded as alternative superconducting planes like the Cu-O planes in cuprates. Theorists suggested that superconductivity may be induced by doping low-spin Ni$^{2+}$ ($S=0$) to a nickelate antiferromagnetic (AF) insulator with Ni$^{1+}$ ($S=1/2$) in a square planar coordination with O ions\cite{anisimov1999electronic}. This could be realized in the chemical reduced RP phase $Ln_{n+1}$Ni$_n$O$_{2n+2}$ by hole doping on the $Ln$ site, such as Sr doped $Ln_{0.8}$Sr$_{0.2}$NiO$_2$, where $Ln$=La, Nd, and Sm. Superconductivity has been indeed observed in films of these hole doped compounds and Nd$_6$Ni$_5$O$_{12}$, where nickel ions exhibit an average valence of +1.2\cite{Gu2022,Pan2022}. The transition temperature $T_c$ of the superconducting films could be enhanced by pressure\cite{Wang2021}. However, superconductivity has not been observed in bulk samples of the $Ln$-Ni-O system under ambient or high pressure\cite{wang2020synthesis,He2021,li2020absence,Huo2022}. 

\begin{table}[b]
\caption{\label{tab:table1}Empirical formula, corresponding $n$, and the average Ni valence of the RP phases $Ln_{n+1}$Ni$_n$O$_{3n+1}$ and the chemical reduced phases $Ln_{n+1}$Ni$_n$O$_{2n+2}$, $Ln$ = La, Pr, Nd, Sm, and Eu. }
\begin{tabular}{c|cc|cc}
\hline \hline
\multirow{2}{*}{n}    & \multicolumn{2}{l|}{$Ln_{n+1}$Ni$_n$O$_{3n+1}$}        & \multicolumn{2}{l}{$Ln_{n+1}$Ni$_n$O$_{2n+2}$} \\ \cline{2-5}  
                      & composition               & Ni valence              & composition               & Ni valence              \\     \hline
$\infty$ & 113                  & +3                      & 112                  & +1                      \\     \hline
3                     & 4310                 & +2.67                   & 438                  & +1.33                   \\     \hline
2                     & 327                  & +2.5                    & 326                  & +1.5                    \\     \hline
1                     & 214                  & +2                      & \multicolumn{2}{c}{-}                      \\     \hline \hline       
\end{tabular}
\end{table}

Progress on studies of the charge and spin order has been made due to the availability of high-quality single crystals for the La-Ni-O system grown by the floating zone technique with high oxygen pressure. AF transitions were identified on metallic LaNiO$_3$ (n=$\infty$) and La$_4$Ni$_3$O$_{10}$ (n=3) single crystals by neutron scattering studies, which were absent for previous measurements on powder samples\cite{guo2018antiferromagnetic,zhang2020intertwined,garcia1992neutron}. The AF transition has been ascribed to spin density wave (SDW) that originates from Fermi surface nesting, differing from the spin order in doped $n=1$ La$_{2-x}$Sr$_x$NiO$_4$ and cuprates. A charge density wave (CDW) coincident with the SDW was found in La$_4$Ni$_3$O$_{10}$ and was suggested to result in a metal-to-metal transition. For the topochemically reduced product La$_4$Ni$_3$O$_8$, synchrotron X-ray and neutron diffraction also reveal stacked charge and spin stripe weakly correlated along the $c$ axis\cite{zhang2016stacked,zhang2019spin}. In this case, La$_4$Ni$_3$O$_8$ is an insulator and the charge and spin order results from the competition between the Coulomb repulsion, spin orbital coupling, and magnetic exchange interaction.

The n=2 RP compound La$_3$Ni$_2$O$_7$ and chemical reduced product La$_3$Ni$_2$O$_6$ consisting of the bilayer NiO$_2$ planes are analogous to the trilayer La$_4$Ni$_3$O$_{10}$ and La$_4$Ni$_3$O$_{8}$ in structural and physical properties\cite{sreedhar1994low,zhang1994synthesis,Taniguchi1995,kobayashi1996transport,Ling1999,zhang1995synthesis}. Theoretical calculations for La$_3$Ni$_2$O$_7$ and La$_4$Ni$_3$O$_{10}$ suggest existence of a hidden one-dimensional Fermi surface nesting which would result in CDW instability\cite{wu2001magnetic,seo1996electronic}. For La$_3$Ni$_2$O$_6$ and La$_4$Ni$_3$O$_{8}$, charge-ordered related structural distortion and magnetic order will emerge in the ground state\cite{botana2016charge}. The electronic density of states of La$_3$Ni$_2$O$_7$ indeed has an abrupt change at around $100\sim120$ K, reflected in both the Hall coefficient and Seebeck coefficient\cite{sreedhar1994low,Taniguchi1995,kobayashi1996transport}. However, neutron diffraction\cite{Ling1999}, resistivity, and magnetic susceptibility measurements on powder samples of La$_3$Ni$_2$O$_7$ that were synthesized through the soft chemistry method did not reveal evidence of charge or spin order\cite{sreedhar1994low,zhang1994synthesis,Taniguchi1995,kobayashi1996transport,Ling1999}. La$_3$Ni$_2$O$_6$ has raised great interest because of the similarities of the electronic structures to cuprates and possible superconductivity through tuning the valence by carrier doping or high pressure\cite{poltavets2009electronic}. Moreover, NMR study of La$_3$Ni$_2$O$_6$ suggests the existence of magnetic interactions\cite{crocker2013nmr}, but the basic magnetic properties are still not clear. As a matter of fact, a comprehensive study of La$_3$Ni$_2$O$_7$ and La$_3$Ni$_2$O$_6$ is lacking due to the unavailability of single crystals. Single crystal growth of the samples with the average valences of nickel ions larger than +2 requires high pressure oxygen and the pressure window is narrow for a specific compound\cite{zhang2020high}. However, single crystals are crucial for investigations of the possible emerging orders of charge and spin in La$_3$Ni$_2$O$_7$ and La$_3$Ni$_2$O$_6$ in order to ascertain the universality for the ternary nickel oxide system, and pave a way for further manipulation of the states of nickel, even for realization of superconductivity\cite{poltavets2009electronic}.

Here, we report successful growth of La$_3$Ni$_2$O$_7$ single crystals by a high oxygen pressure floating zone technique and the achievement of La$_3$Ni$_2$O$_6$ single crystals through a subsequent low temperature topochemical reduction method\cite{zhang2020high}. Electrical resistivity measurements on the single crystals of La$_3$Ni$_2$O$_7$ and La$_3$Ni$_2$O$_6$ show significantly different properties. La$_3$Ni$_2$O$_7$ is metallic while La$_3$Ni$_2$O$_6$ is insulating. Superconductivity has not been realized in La$_3$Ni$_2$O$_6$ under pressure up to 25.3 GPa. Anomalies in resistivity, susceptibility, and specific heat that may correspond to spin and charge order have been observed in both compounds. The results indicate the emergent order of charge and spin is universal for the nickelates with mixed-valent states of nickel.

\section{Experimental Methods}

Polycrystalline samples were synthesized through the standard solid-state reaction techniques. Stoichiometric amounts of La$_2$O$_3$ and  an excess of 0.5\% NiO (Macklin, 99.99\%) were thoroughly ground. The excess of 0.5\% NiO was used to compensate the loss of volatilization. The ground mixtures were made into pellets and sintered at 1100 $^\circ$C for 50 h. After cooling down to room temperature, the reactants were reground and made into pellets for sintering again. The procedures were repeated for 3 times to ensure a complete and homogeneous reaction\cite{zhang2020high}. The seed and feed rods were prepared by pressuring the powders under hydrostatic pressure and sintered at 1400 $^\circ$C for 12 h. The rods were approximate 9 cm in length and 0.6 cm in diameter.

High oxygen pressure is crucial for synthesis of the homologous RP system of nickelates. A vertical optical-image floating-zone furnace designed for a 100 bar high pressure (HKZ, SciDre, Dresden) was employed in our single crystal growth. La$_3$Ni$_2$O$_7$ single crystals were grown with an oxygen pressure at p(O$_2$)=15 bar and a 5 kW Xenon arc lamp. The traveling rate was 3 mm/h after a fast traveling procedure at 10 mm/h  to improve the density. After that, the feed and seed rods counter-rotate at 15 and 10 rpm, respectively, in order to improve homogeneity. La$_3$Ni$_2$O$_6$ was obtained from the La$_3$Ni$_2$O$_7$ single crystals by the topochemical reduction method. La$_3$Ni$_2$O$_7$ single crystals were enclosed in an aluminum foil and sealed in vacuum with stoichiometric CaH$_2$ powders. The reaction was under the condition of 300 $^\circ$C for 4 days.

X-ray photoemission spectroscopy (XPS) measurements was carried out on an XPS machine (Escalab 250 Xi, Thermo Fisher). Argon sputtering was adopted to remove the surface contamination. The sputtering depth was about 100 nm. The monochromatic Al K$_{\alpha}$ radiation with photon energy of 1486.6 eV was applied to analyses the valent states of nickel. The X-ray beam was focused on a 0.5 mm spot surface. In order to obtain high-resolution spectra, the electron energy analyzer was operated at a pass energy of 30 eV. The C 1$s$ photoelectron line (284.8 eV) was used to calibrate the binding energies of the photoelectrons.

Magnetic susceptibility, resistivity, and specific heat measurements were performed on a physical property measurement system (PPMS, Quantum Design). \emph{In situ}  high-pressure electrical resistance measurements were carried out in a diamond anvil cell made from a Be-Cu alloy using a standard four-probe technique\cite{cai2020,sun2021}. NaCl powders were employed as the pressure transmitting medium. The pressure in the resistance measurements was calibrated by the ruby fluorescence shift at room temperature. X-ray single crystal diffraction (XRD) was performed on a single-crystal X-ray diffractometer ( SuperNova, Rigaku) using the Mo-K$_\alpha$ radiation at 300 K. The diffraction data were refined by the Rietveld method\cite{Rietveld1969}. Energy-dispersive X-ray spectroscopy (EDS) (EVO, Zeiss) was employed to determine the compositions of the crystals. In addition, Laue X-ray diffraction technique was utilized to confirm the crystal orientation and single crystallinity. 

\begin{table*}[t]
	\caption{\label{tab:table2} Structural parameters for single crystals of La$_3$Ni$_2$O$_7$ and La$_3$Ni$_2$O$_6$ at 300 K measured with the Mo K$_{\alpha}$ radiation with $\lambda=0.7107$ \AA. The data were collected in the range of $1.171^{\circ}\le2\theta \le40.84^{\circ}$. }
	\begin{ruledtabular}
		\begin{tabular}{ccc}
			\textrm{Empirical formula}&\multicolumn{1}{c}{La$_3$Ni$_2$O$_7$}&\multicolumn{1}{c}{La$_3$Ni$_2$O$_6$}\\ \hline
			Space group&Cmcm (Orthorhombic)&I4/mmm (Tetragonal)\\
			Unit-cell parameters   &  $a=5.4000(7)$ \AA; $b=5.4384(7)$ \AA &$a=b=3.9720(4)$ \AA\\
			&$c=20.455(4)$ \AA   &$c=19.368(4)$ \AA         \\
			& $\alpha=\beta=\gamma=90^{\circ}$     &$\alpha=\beta=\gamma=90^{\circ}$\\
			\hline
			
			La1&$(0.32,0.75,0.25)$&$(1,1,0.5)$\\
			La2&$(0.5,0.25,0.75)$&$(1,0,0.32)$\\
			Ni&$(0.40,0.75,0.75)$&$(0.5,0.5,0.42)$\\
			O1&$(0.41,1,1)$&$(1,0.5,0.42)$\\
			O2&$(0.5,0.71,0.75)$&$(0.5,0,0.25)$\\
			O3&$(0.40,0.5,0.5)$&$-$\\
			O4&$(0.30,0.78,0.75)$&$-$\\

			\hline
			Goodness-of-fit on $F^2$&1.184& 0.966\\
			Final $R$ indexes (all data)&$R_1$=0.0327&$R_1$=0.0469\\
		\end{tabular}
	\end{ruledtabular}
\end{table*}

\section{Results}

\subsection{La$_3$Ni$_2$O$_7$}

\begin{figure}[b]
        \includegraphics[scale=0.23]{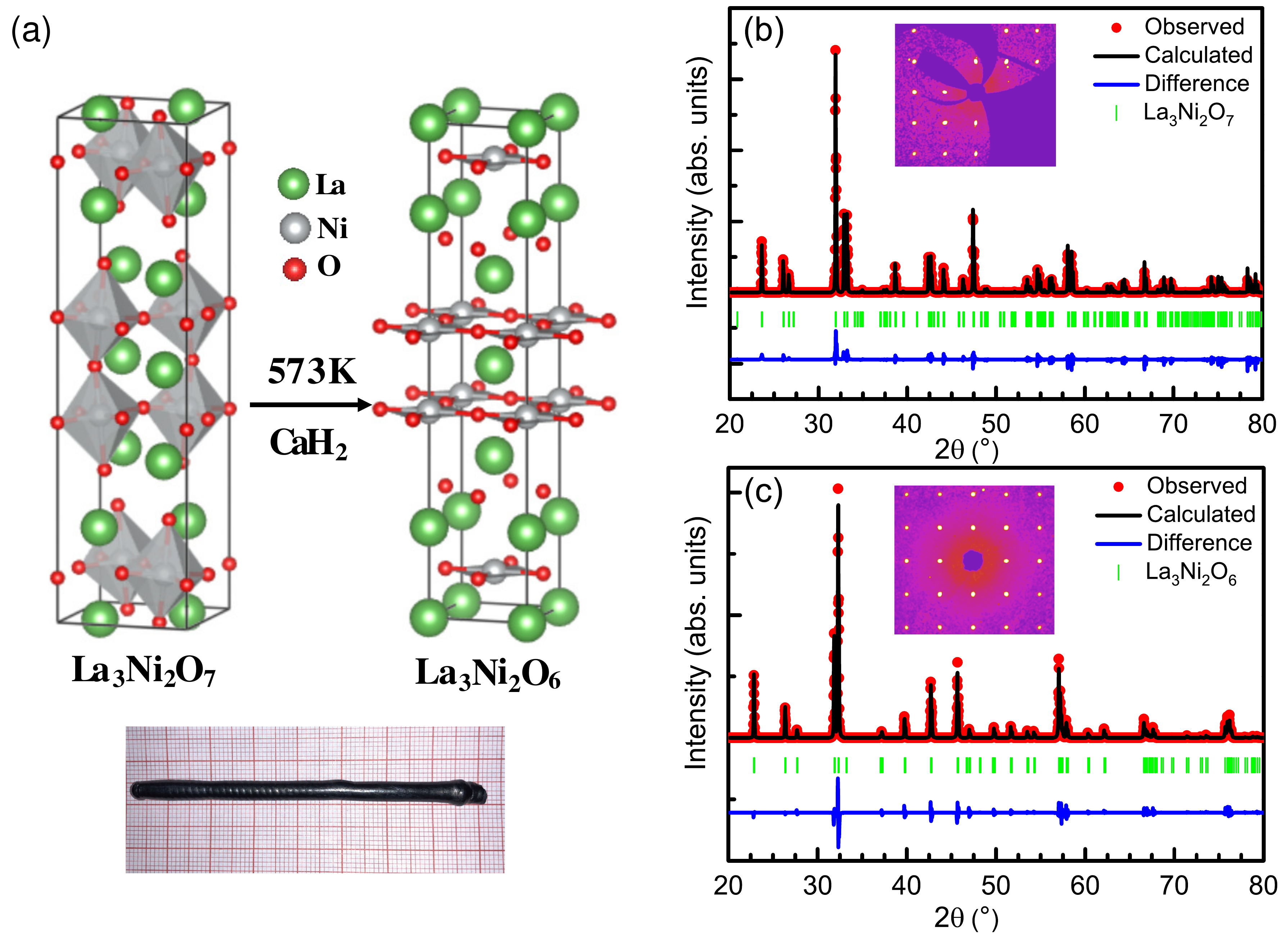}
	\caption{(a) Sketches of the crystal structures of La$_3$Ni$_2$O$_7$ and La$_3$Ni$_2$O$_6$. A photo of the single crystal of La$_3$Ni$_2$O$_7$. (b) X-ray diffraction patterns integrated from single crystal diffraction on La$_3$Ni$_2$O$_7$ and (c) La$_3$Ni$_2$O$_6$. The insets in (b) and (c) are diffraction patterns in reciprocal space on corresponding single crystals collected on a single crystal X-ray diffractometer.}
	\label{fig1}
\end{figure}

Crystal structures of La$_3$Ni$_2$O$_7$ and La$_3$Ni$_2$O$_6$ are shown in Fig. \ref{fig1}(a). La$_3$Ni$_2$O$_7$ crystallizes in orthorhombic symmetry (space group: Cmcm) with distorted vertex-sharing NiO$_6$ octahedra and rock-salt La-O layers\cite{Ling1999}. The structure can be termed as inter-growth of two NiO$_6$ octahedra planes and a La-O fluorite-type layer, stacking along the $c$ direction. La$_3$Ni$_2$O$_6$ with the apical oxygen atoms removed crystallizes in tetragonal symmetry (space group: I4/mmm). The structure is stacked by two corner-sharing square NiO$_2$ planes and a La-O fluorite-type layer along the $c$ direction\cite{poltavets2006la3ni2o6}. The XRD patterns measured on single crystals reveal high quality of our samples [Fig. \ref{fig1}(b) and  \ref{fig1}(c)]. Detailed Rietveld-refined structural results are summarized in Table \ref {tab:table2}. The compositions of the single crystals determined by EDS are La$_{2.95}$Ni$_{2}$O$_7$ and La$_{2.92}$Ni$_{2}$O$_6$, respectively, close to the stoichiometric compositions within the instrumental accuracy. We note the oxygen content is not sensitive in EDS and the contents of La have been normalized by that of Ni.

\begin{figure}[b]
	\centering
	        \includegraphics[scale=0.35]{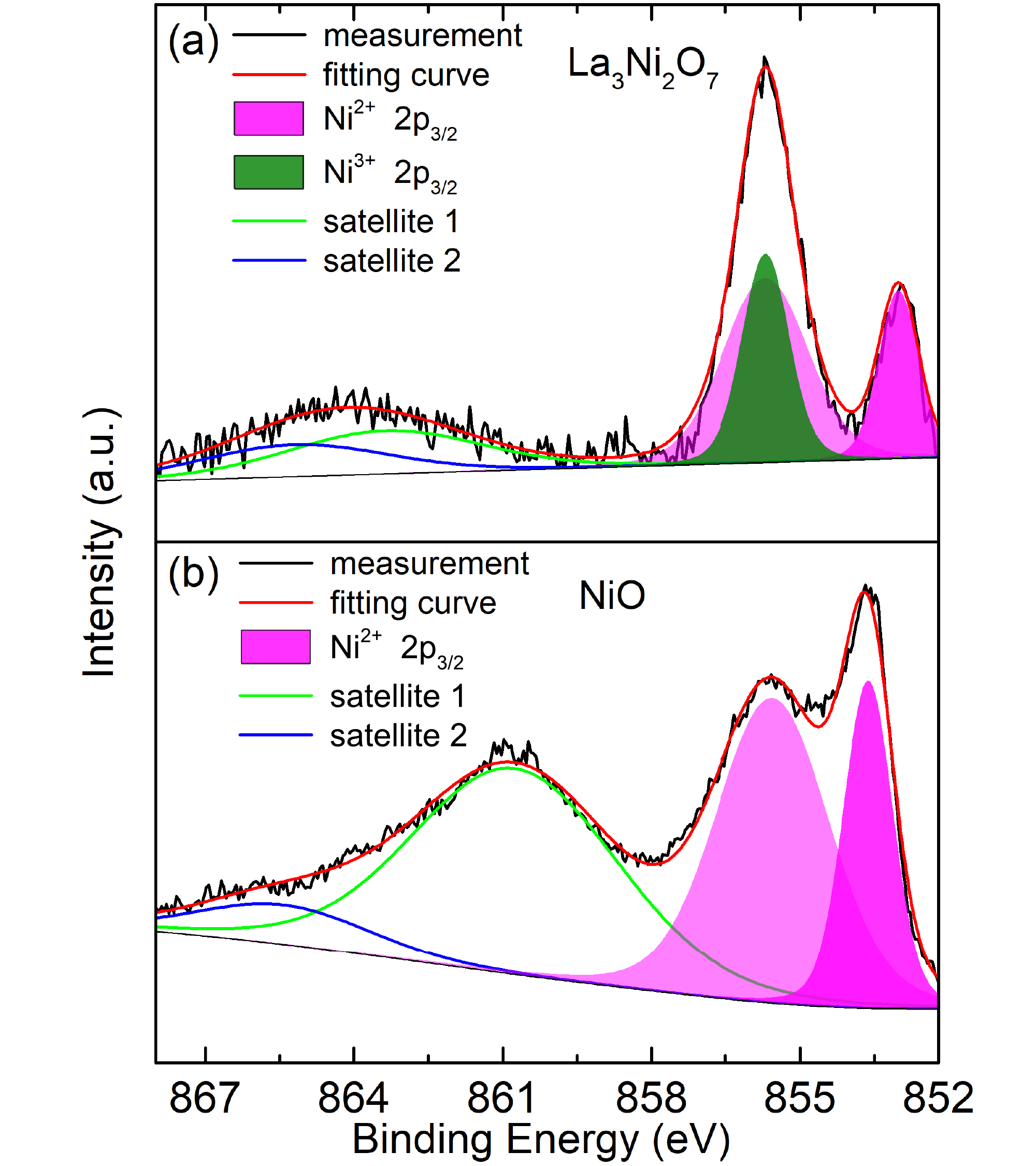}
	\caption{(a) XPS measurements of the Ni 2$p$ core levels for La$_3$Ni$_2$O$_7$ and (b) NiO. The pink areas and satellites 1 and 2 in (b) are deconvoluted of the Ni 2$p_{3/2}$ XPS data. The ratio of the two main peaks for Ni$^{2+}$ are preserved in the fitting of the Ni 2$p$ core levels for La$_3$Ni$_2$O$_7$ in (a). The black solid lines are the experimental data. The red line is a fitting of the total intensities.} 
		\label{fig2}
\end{figure}

The valent states of Ni are crucial for realization of charge and spin order. La$_3$Ni$_2$O$_7$ is metallic, where nickel ions could host a valence of +2.5, or a mixed-valent states of +2 and +3. Figure \ref{fig2} (a) shows the XPS spectrum of Ni ions in La$_3$Ni$_2$O$_7$. As a comparison, an XPS spectrum of Ni$^{2+}$ measured on NiO is presented in Fig. \ref{fig2} (b). The main peaks of the spectrum of Ni$^{2+}$ locate at 853.6 and 855.6 eV. In addition, a broad satellite peak appears at a higher binding energy of 860.8 eV. These features are typical for nickel ions with the divalent oxidation state. The XPS spectrum of La$_3$Ni$_2$O$_7$ exhibits asymmetric doublet peaks at 853.0 and 855.7 eV as shown in Fig. \ref{fig2} (a). We preserve the ratio of the two peaks of Ni$^{2+}$ as measured on NiO and fit the spectrum of La$_3$Ni$_2$O$_7$. A peak at 855.7 eV could be separated, yielding the existence of the trivalent oxidation state\cite{Qiao2011,Liu2019,Chen2019}. Thus, our XPS measurements reveal that the valent states of nickel in La$_3$Ni$_2$O$_7$ are a mixture of the divalent Ni$^{2+}$ and trivalent Ni$^{3+}$ oxidation states.

\begin{figure}[t]
	\centering
               \includegraphics[scale=0.25]{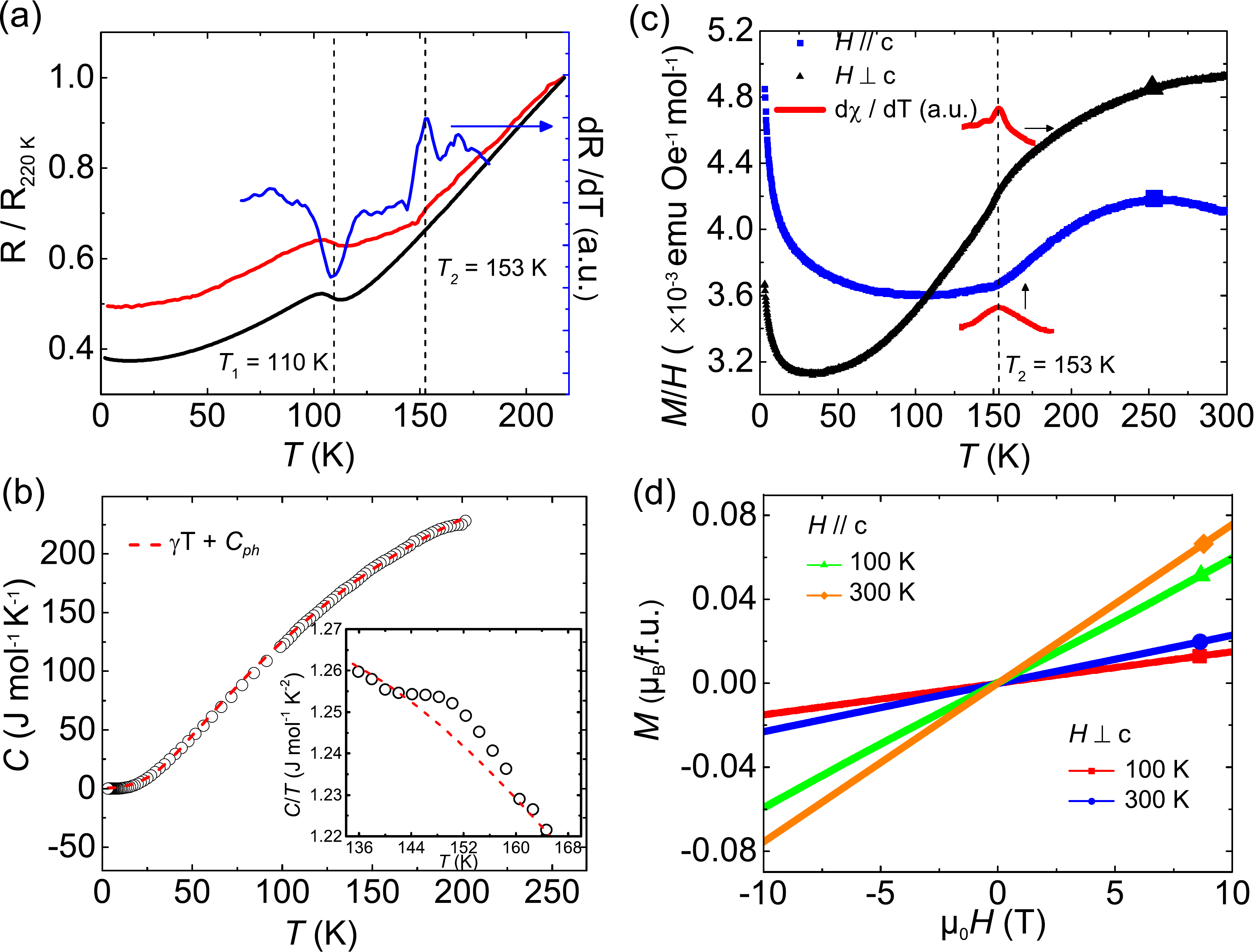}
	\caption{
	(a) The red and black lines are resistance of two La$_3$Ni$_2$O$_7$ single crystals as a function of temperature. The blue line is a derivative of the resistance in red. (b) Temperature dependence of the total specific heat of La$_3$Ni$_2$O$_7$. The dashed line is a fitting to the formula described in the text. The inset shows an abnormal change at $\sim$153 K. (c) Magnetization of La$_3$Ni$_2$O$_7$ measured with a magnetic field of $\mu_0H=0.4$ T parallel to and perpendicular with the $c$ axis under zero-field cooled condition. The redlines show derivatives of the magnetization in arbitrary units. (d) Magnetization against the magnetic field $\mu_0H$ for $H\parallel c$ and $H\perp c$ at 100, and 300 K.
	}
	\label{fig3}
\end{figure}

Figure \ref{fig3}(a) shows the temperature dependence of the resistance for La$_3$Ni$_2$O$_7$ single crystals, revealing a metallic ground state that is analogous to the $n=3$, and $\infty$ RP compounds\cite{zhang1994synthesis,guo2018antiferromagnetic,rivadulla2003electron,liu2020observation}. We observe two anomalies in resistance, one at $\sim$110 K and the other one at 153 K. They could be identified unambiguously in the deviation of resistance against temperature $dR/dT$ as shown in Fig. \ref{fig3} (a). The former one at $\sim$ 110 K is close to the anomalies observed in Hall and Seebeck coefficients\cite{sreedhar1994low,Taniguchi1995,kobayashi1996transport}. While the later one at 153 K is weaker and has not be identified in the previous measurements on powder samples. For LaNiO$_3$ and La$_4$Ni$_3$O$_{10}$, a metal-metal transition near 150 K has been observed in resistivity, where the origin has been proved to result from intertwined charge and spin density wave\cite{guo2018antiferromagnetic,zhang2020intertwined}. 

To investigate the origin of the transitions in La$_3$Ni$_2$O$_7$, we measured the specific heat and magnetization against temperature and magnetic field. The specific heat for temperatures between 3 and 200 K is shown in Fig. \ref{fig3} (b). A model \textit{C} = \textit{$\gamma$T} + \textit{C$_{ph}$} was employed to fit the data, where \textit{$\gamma$T} and \textit{C$_{ph}$} represent the contributions of electrons and phonons, respectively. A modified Debye model considering the existence of two phonon modes that reconcile the heavy atoms (La and Ni) and light atoms (O) was considered to describe the phonon contribution, $C_{ph}=9R\sum\limits_{n=1}^{2}C_n(\dfrac{T}{\theta_{Dn}})^{3}\int_{0}^{\theta_{Dn}/T}\dfrac{x^4e^x}{(e^x-1)^2}dx$, where $R = 8.314$ Jmol$^{-1}$K$^{-1}$ is the ideal gas constant and \textit{C$_n$} represents the numbers of the heavy or light atoms in a formula unit. The modeling reveals that, of the 12 atoms in the formula unit, 5 atoms have a Debye temperature \textit{$\theta_{D1}$} of $298\pm3$ K and 7 atoms have a Debye temperature \textit{$\theta_{D2}$} of $620\pm4$ K\cite{li2020magnetic,ortega2006magnetic}.The fitting also yields $\gamma$ = 7.3 mJmol$^{-1}$K$^{-2}$, close to the value of 6.4 mJmol$^{-1}$K$^{-2}$ revealed from powder samples\cite{wu2001magnetic}. With the estimated densities of charge-carriers and mass of the free-electrons, the value of electron effective mass m$^*$/m$_0$ of La$_3$Ni$_2$O$_7$ is $\approx$ 2.12, reminiscent of 2.56 for La$_4$Ni$_3$O$_{10}$ and much lower than 15 for LaNiO$_3$\cite{rajeev1991low,sreedhar1994low,zhang1995synthesis}. The effective mass of electrons suggests that the electronic correlations in La$_3$Ni$_2$O$_7$ and La$_4$Ni$_3$O$_{10}$ are weaker than that of the n=$\infty$ compound LaNiO$_3$. The anomaly at 153 K could also be observed in specific heat, as shown in the inset of Fig. \ref{fig3} (b), in consistence with the temperature of the anomaly at 153 K in resistance.

Figure \ref{fig3} (c) displays temperature dependence of the magnetization of La$_3$Ni$_2$O$_7$ single crystals at 4000 Oe. The kink at $\sim$153 K for both  $H\parallel c$ and $H\bot c$ could be identified in magnetization and $d\chi/dT$, suggesting a same physical origin as the anomalies at the identical temperature in resistance and specific heat. While the anomaly at $\sim$110 K in resistance does not show in magnetization. The upturn below 50 K with decreasing temperature in Fig. \ref{fig3} (c) may be related to magnetic impurities or lattice imperfections. We note the magnetization of La$_3$Ni$_2$O$_7$ single crystals shown in Fig. \ref{fig3} (c) is reminiscent of that observed in LaNiO$_3$ and La$_4$Ni$_3$O$_{10}$ single crystals\cite{guo2018antiferromagnetic,zhang2020intertwined}. For the later two materials, intertwined charge and spin density wave has been confirmed by neutron and X-ray diffraction measurements. The magnetization as a function of magnetic field and temperature is shown in Fig. \ref{fig3} (d). The magnetization evolutes linearly against magnetic field up to $\mu_0H=10$ T, indicating antiferromagnetic correlations. In the scenario of spin order, the anisotropy of the magnetization in Fig. \ref{fig3} (d) suggests the moment of Ni aligned in-plane.

\subsection{La$_3$Ni$_2$O$_6$}

\begin{figure}[b]
	\centering
	\includegraphics[scale=0.3]{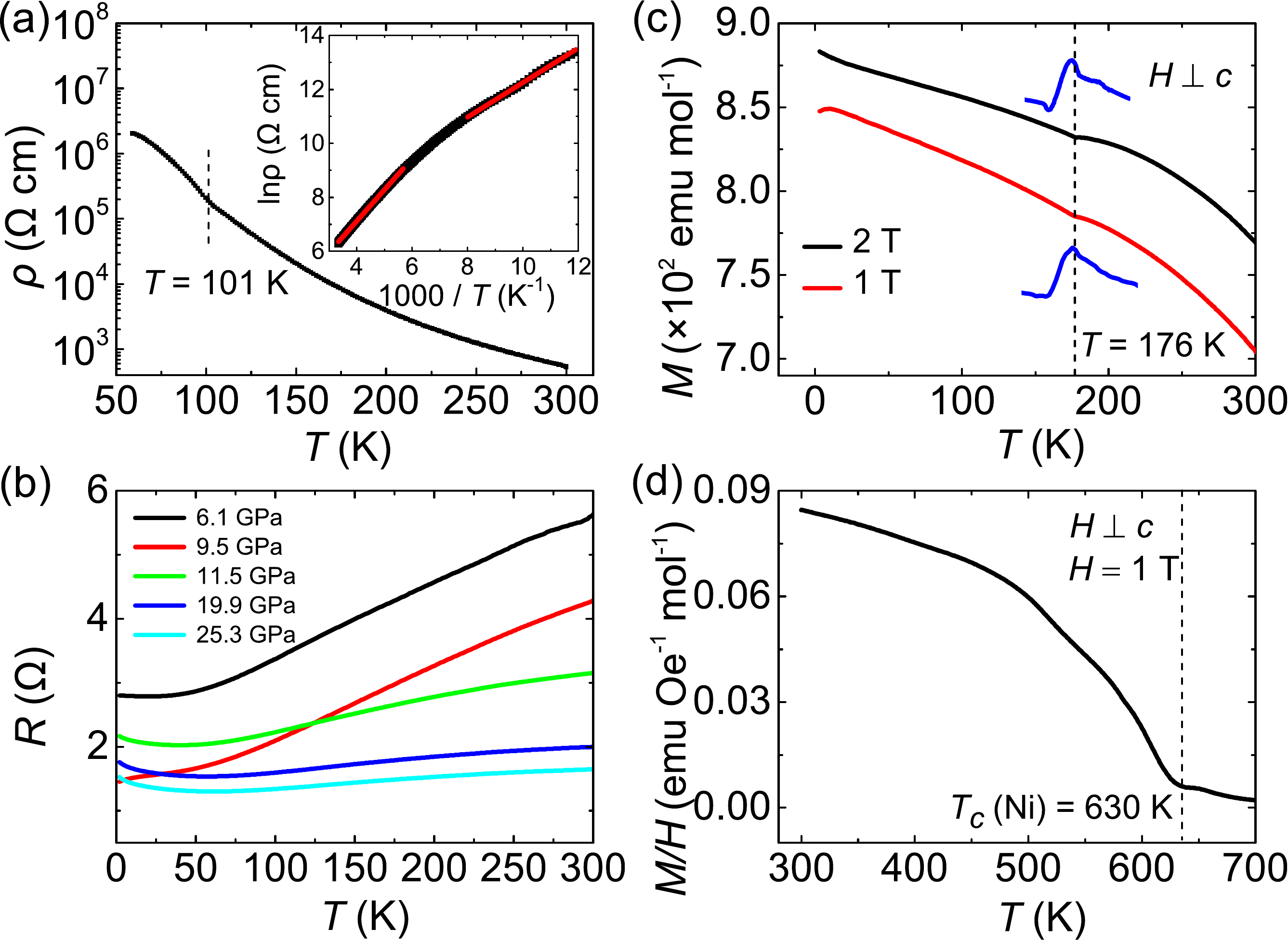}
	\caption{ (a) Resistivity of La$_3$Ni$_2$O$_6$ measured as a function of temperature. The dashed line indicates a weak kink at 101 K. The inset shows a fit using $\rho(T)=\rho_0$exp$(E_a/k_BT)$, resulting in $E_a=54$ meV for $84\le T\le125$ K and $E_a=100$ meV for $176\le T\le300$ K.  (b) Resistance at various pressures applied using a diamond anvil cell. (c) Magnetization measured with magnetic fields of $\mu_0H=1$ and 2 T and perpendicular to the $c$ axis. The blue lines show derivatives of the magnetization. The dashed line indicates a kink at 176 K. (d) High-temperature magnetization measured with $\mu_0H=1$ T and $H\perp c$. The dashed line indicates the ferromagnetic transition of Ni at 630 K.}
	\label{fig4} 
\end{figure}

The ground state of La$_3$Ni$_2$O$_6$ with mixed Ni$^{1+} d^9 (S=1/2)$ and Ni$^{2+} d^8 (S=0)$ was predicted to be an AF insulator with a checkerboard charge order and both the AF and ferromagnetic (FM) interactions. Pressure will suppress the FM interactions further and may induce superconductivity\cite{botana2016charge}. However, only was the insulating state confirmed due to the limitation of powder samples\cite{poltavets2009electronic}. We obtain single crystals of La$_3$Ni$_2$O$_6$ from topochemical reduction of the La$_3$Ni$_2$O$_7$ single crystals. Temperature dependence of the electrical resistivity \textit{$\rho$} is shown in Fig. \ref{fig4} (a) and the inset. The semiconducting behavior is analogous to the measurements on powder samples\cite{poltavets2009electronic}. The resistivity of La$_3$Ni$_2$O$_6$ displays a weak kink at 101 K, in contrast to the abrupt change in resistivity at $\sim$100 K for La$_4$Ni$_3$O$_8$ that corresponds to the formation of charge and spin stripe order\cite{zhang2016stacked,zhang2019spin}. By fitting the resistivity to the activation-energy model $\rho(T)=\rho_0$exp$(E_a/k_BT)$, two thermal activation energy gaps of $E_a=54$ and 100 meV are obtained, corresponding to $84\le T\le125$ K and $176\le T\le300$ K, respectively. To explore superconductivity in La$_3$Ni$_2$O$_6$, the resistance measured under pressure is shown in Fig. \ref{fig4} (b). The metallization is achieved at 6.1 GPa. However, no superconductivity emerges in our samples up to 25.3 GPa.

Figure \ref{fig4} (c) displays the magnetization of La$_3$Ni$_2$O$_6$ with the magnetic field perpendicular to the $c$ axis. A kink at $\sim$176 K could be identified in the derivative of magnetization for the magnetic fields at 1 and 2 T. As theoretical suggestions, the kink on magnetization at 176 K may be associated with the charge and spin order of La$_3$Ni$_2$O$_6$\cite{botana2016charge}, which may also result in the change of the fitted thermal activation gaps against temperature as shown in the inset of Fig. \ref{fig4} (a). The magnetization decreases as the temperature increasing, similar to the behavior of a ferromagnet. It is known that the topochemical reduction method would induce Ni impurity\cite{li2020absence}. A high-temperature magnetization measurement was conducted, revealing a FM transition at $T_c=630$ K, as shown in Fig. \ref{fig4} (d). The result demonstrates the existence of Ni impurity.

\section{Discussion and Summary}    

The mixed-valent and spin states of Ni in the ternary La-Ni-O system tend to form charge and spin order. The emergence of the two types of order has been verified in hole doped La$_{2-x}$Sr$_x$NiO$_4$ ($x=$1/4, 1/3, and 1/2), La$_4$Ni$_3$O$_{10}$, and La$_4$Ni$_3$O$_{8}$\cite{Chen1993,Lee1997,Zhong2017,zhang2016stacked,zhang2019spin}.
For the trilayer RP compounds, the ratio of the magnetic Ni$^{3+}$ ($S=1/2$) and nonmagnetic Ni$^{2+}$ ($S=0$) is $2:1$; for the bilayer RP compounds, the value of this ratio drops to $1:1$. The formation of charge order is expected, while the spin order may be weaker in the bilayer compounds. The changes in resistivity, susceptibility, and specific heat at 153 K are indeed less pronounced in La$_3$Ni$_2$O$_7$ single crystals compared to that of La$_4$Ni$_3$O$_{10}$. The possibility that the charge order emerges in the bilayer La-Ni-O system without accompanying by the spin order could not be ruled out. In this case, the anomaly in magnetization could be due to charge-spin interactions. Our Raman scattering measurements on La$_3$Ni$_2$O$_7$ also reveal an anomaly at $\sim$150 K for the position of the peak at $\sim$597 cm$^{-1}$ (data not shown). Based on previous studies, the anomaly in resistance at $\sim$110 K may be related to the change of carrier concentration induced by a structural evolution against temperature\cite{sreedhar1994low,zhang1994synthesis,Taniguchi1995,kobayashi1996transport,Ling1999}. 

For La$_3$Ni$_2$O$_6$, an insulating ground state with a checkerboard charge order and AF order based on Ni$^{1+}$ is expected\cite{botana2016charge}. The average valence of Ni, +1.5, is close to +1.2 that of the superconducting film samples. Evidences for the charge and spin order have been observed. However, superconductivity does not appear up to 25.3 GPa, where the samples have been metallized\cite{poltavets2009electronic}. The superconductivity seems to be sensitive to the average-valent state of nickel. In addition, the possible charge and spin order in La$_3$Ni$_2$O$_6$ may suppress superconductivity under pressure.
Further synchrotron X-ray and neutron diffraction studies on single crystal samples of the bilayer compounds are necessary to demonstrate the charge and spin order.

In summary, we have grown the bilayer nickelates La$_3$Ni$_2$O$_7$ and La$_3$Ni$_2$O$_6$ single crystals successfully by the floating-zone method with high pressure of oxygen. The structural, magnetic, electronic, and specific heat properties of both compounds are characterized in detail. Electronic measurements show metallic property for La$_3$Ni$_2$O$_7$, and semiconducting for La$_3$Ni$_2$O$_6$ with a thermal activation gap of 55 meV. The resistance, magnetization, and specific heat all reveal a transition-like anomaly at 153 K for La$_3$Ni$_2$O$_7$, suggesting the formation of charge and spin order. In addition, the magnetization of La$_3$Ni$_2$O$_6$ also yields a kink at 176 K, reminiscent to the charge and spin order in La$_3$Ni$_2$O$_7$. Pressure above 6.1 GPa could metallize La$_3$Ni$_2$O$_6$. However, no superconductivity is observed up to 25.3 GPa. Our data suggest the formation of charge and spin order may be a universal characteristic for nickelates with mixed-valent states of nickel.

\section{Acknowledgments}

Work was supported by the National Natural Science Foundation of China (Grants No. 12174454, 11904414, 11904416, U2130101), the Guangdong Basic and Applied Basic Research Foundation (No. 2021B1515120015), and National Key Research and Development Program of China (No. 2019YFA0705702, 2020YFA0406003, and 2021YFA1400401, 2021YFA0718900).

\bibliography{reference}
\end{document}